\documentclass[reprint,floatfix,notitlepage,nofootinbib,twocolumn,superscriptaddress,pra,preprintnumbers]{revtex4-1}
\usepackage{amsfonts,amsmath,amssymb, amsthm}
\usepackage{graphicx,tikz,float,caption,subcaption} % Required for inserting images
\usepackage{mathrsfs}
\usepackage{enumitem}
\usepackage{pifont}
\usepackage{algorithm}
\usepackage{algpseudocode}
\usepackage[english]{babel} 
\usepackage[autostyle, english = american]{csquotes}

\makeatletter
\def\bbl@set@language#1{%
  \edef\languagename{%
    \ifnum\escapechar=\expandafter`\string#1\@empty
    \else\string#1\@empty\fi}%
  \@ifundefined{babel@language@alias@\languagename}{}{%
    \edef\languagename{\@nameuse{babel@language@alias@\languagename}}%
  }%
  \select@language{\languagename}%
  \expandafter\ifx\csname date\languagename\endcsname\relax\else
    \if@filesw
      \protected@write\@auxout{}{\string\select@language{\languagename}}%
      \bbl@for\bbl@tempa\BabelContentsFiles{%
        \addtocontents{\bbl@tempa}{\xstring\select@language{\languagename}}}%
      \bbl@usehooks{write}{}%
    \fi
  \fi}
\newcommand{\DeclareLanguageAlias}[2]{%
  \global\@namedef{babel@language@alias@#1}{#2}%
}
\makeatother
\DeclareLanguageAlias{en}{english}

\makeatletter
\renewcommand{\fnum@algorithm}{\fname@algorithm~\thealgorithm}
\makeatother

\usepackage{ragged2e,etoolbox}

\usepackage{bm}
\renewcommand{\vec}[1]{\boldsymbol{\mathbf{#1}}}

\usepackage[colorlinks=true,citecolor=blue,linkcolor=magenta]{hyperref}

\graphicspath{{./}{./images/}}

\newtheorem{definition}{Definition}
\newtheorem{theorem}{Theorem}
\DeclareMathOperator{\Tr}{Tr}
\DeclareMathOperator{\tr}{tr}
\newcommand{\ket}[1]{|#1\rangle}
\newcommand{\bra}[1]{\langle #1|}
\newcommand{\poly}[0]{\text{Poly}}
\newcommand{\M}[0]{\mathcal{M}}
\newcommand{\mH}[0]{\mathcal{H}}
\newcommand{\mS}[0]{\mathcal{S}}
\newcommand{\stab}[0]{\text{STAB}}

\newcommand{\rank}[0]{\text{rank}}
\newcommand{\N}[0]{\text{N}}

\newcommand{\lrp}[1]{\left(#1\right)}

\begin{document}

%\title{An efficient holographic algorithm to compute entanglement in states with low magic}
\title{An efficient algorithm to compute entanglement in states with low magic}
\author{ChunJun Cao}
\author{Gong Cheng}
\author{Tianci Zhou}
\affiliation{Department of Physics, Virginia Tech, Blacksburg, VA, USA 24061}
\affiliation{Virginia Tech Center for Quantum Information Science and Engineering, Blacksburg, VA 24061, USA}

\date{\today}

\begin{abstract}
%The Ryu-Takayanagi formula plays an important role in holography. A generalized form of it holds for quantum error correcting codes that satisfy complementary recovery which allows one to separate the entanglement entropy of a system into the ``bulk'' and ``area'' entropy contributions. We show how this separation can be combined with the stabilizer formalism to efficiently measure or compute the von Neumann and R\'enyi entropies of states for which no efficient techniques exist. These states include low magic states as well as states that are prepared by entangling the product of magic states by any Clifford unitary, the latter of which can have both reasonably high magic and entanglement.
A bottleneck for analyzing the interplay between magic and entanglement is the computation of these quantities in highly entangled quantum many-body magic states. Efficient extraction of entanglement can also inform our understanding of dynamical quantum processes such as measurement-induced phase transition and approximate unitary designs. We develop an efficient classical algorithm to compute the von Neumann entropy and entanglement spectrum for such states under the condition that they have low stabilizer nullity.  The algorithm exploits the property of stabilizer codes to separate entanglement into two pieces: one generated by the common stabilizer group and the other from the logical state. The low-nullity constraint ensures both pieces can be computed efficiently. Our algorithm can be applied to study the entanglement in sparsely $T$-doped circuits with possible Pauli measurements as well as certain classes of states that have both high entanglement and magic. Combining with stabilizer learning subroutines, it also enables the efficient learning of von Neumann entropies for low-nullity states prepared on quantum devices. 
\end{abstract}

\maketitle
\section{Introduction}
In the era of quantum information, entanglement serves as a versatile probe to understand many-body physics \cite{Nielsen_Chuang_2010,bennett_teleporting_1993,osterloh_scaling_2002,vidal_entanglement_2003,calabrese_entanglement_2004,kitaev_topological_2006,levin_detecting_2006,eisert_area_2010,skinner_measurement-induced_2019}. But in general, it is believed to be a challenge to compute the entanglement of a generic state, as the resources to merely describe the state grow exponentially with the system size $n$. Even sophisticated approaches, such as matrix product states\cite{fannes_matrix_1992,ostlund_thermodynamic_1995,verstraete_matrix_2004,perez-garcia_matrix_2007,schollwock_density-matrix_2011},
have been developed to efficiently capture the entanglement structure of one-dimensional many-body systems, struggle with states with a volume-law entanglement. However, the stabilizer state represents a well-known exception\cite{Nielsen_Chuang_2010,Gottesman1998TheHR,aaronson_improved_2004,gottesman_stabilizer_1997}. Instead of recording the wavefunction components, one only needs to track the change of $n$ generators of its stabilizer group elements so that tasks computing correlation functions and entanglement reduce to the operations of those stabilizer generators \cite{Gottesman1998TheHR,aaronson_improved_2004}, which can be captured by row operations in binary symplectic matrices with computational complexity only polynomial in $n$. Despite the efficiency, stabilizer states are not universal for quantum computing \cite{bravyi_universal_2005,aaronson_improved_2004,Gottesman1998TheHR}. In particular, their entanglement spectra (the singular value spectra of the reduced density matrix) are entirely flat. Many many-body states of interest, both in and out of equilibrium, are usually superpositions of stabilizer states \cite{bravyi_classical_2016,veitch_resource_2014}. This leads to the natural question: Can we compute the entanglement of states that lie just beyond the stabilizer manifold? 

More concretely, consider a quantum state of the form $|\Psi \rangle = \sum_{j=1}^K c_j |\psi_j \rangle$ \cite{bravyi_trading_2016,bravyi_simulation_2019}, where each $|\psi_j \rangle$ is a stabilizer state. As it deviates from a pure stabilizer state, the state $|\Psi \rangle$  acquires ``magic'', a resource for universal fault-tolerant quantum computation and a measure of quantumness that is distinct from entanglement. The sets of stabilizers for these $K$ states, along with the linear coefficients $c_j$, give an efficient description of $|\Psi \rangle$ that is polynomial in the system size. Given a bipartition $A$ and $A^c$ of the system and a pure state $|\Psi \rangle$ in this form, how can we compute the von Neumann entropy for region $A$? More generally, can we efficiently determine the R\'enyi entropy of arbitrary index for this reduced state given that the amount of magic is low?

In this work, we provide an affirmative answer for a subclass of states $|\Psi \rangle$ where an efficient classical algorithm exists. This subclass requires that the constituent states $|\psi_j \rangle$ share a substantial number of common stabilizers, specifically $n - \nu \sim n$ where the integer $\nu$, known as the stabilizer nullity, is a measure of magic \cite{beverland_lower_2020}. The computational complexity of our algorithm is polynomial in $n$ and exponential in $\nu$. Hence our algorithm remains efficient for states with a low stabilizer nullity up to logarithmic in the system size, i.e., $\nu \sim \log(n)$.

Our approach exploits the stabilizer code structure of low-nullity states, which allows us to separate the entanglement into two additive pieces. Intuitively, one piece comes from the stabilizer entanglement associated with the $n - \nu$ common stabilizers shared by the codewords $|\psi_j\rangle$, making this entropy easy to compute. The other component comes from the $\nu$ qubits of logical information within this code. By isolating the logical state, which is of manageable size ($\nu \sim \ln(n)$), we can compute this piece by a brute-force approach.

\begin{figure}
    \centering
    \includegraphics[width=0.9\linewidth]{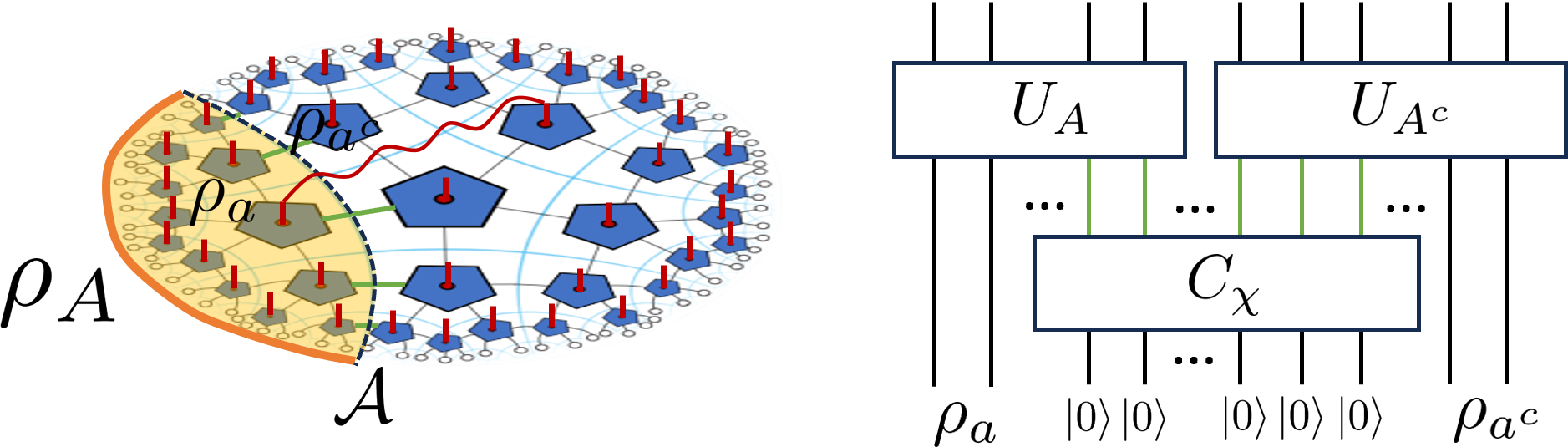}
    \caption{\justifying A holographic stabilizer code where red edges represent bulk logical qubits while the white circles represent physical qubits. All the stabilizer entanglement contributing to $S(\rho_A)$ comes from the minimal surface (dashed curve) area $\mathscr{A}$ captured by the in-plane edges (green). The bulk state can also be entangled across the $k$ logical qubits and only this entanglement can contain non-stabilizerness. The bulk logical qubits in the yellow wedge are in the state $\rho_a$ and are recoverable from subsystem $A$. The remaining bulk qubits are recoverable from $A^c$. Such a code can be built from a circuit on the right where $C_\chi$ creates the entanglement resource $|\chi\rangle$ (on the green wires) that facilitates the entanglement in green on the left.}
    \label{fig:holo}
\end{figure}

Surprisingly, the intuition behind this separation of entanglement is best understood from the point of view of holographic duality \cite{Chen_2022}. %into logical contributions and a common component also emerges in holographic theory. 
The celebrated Ryu-Takayanagi (RT) formula \cite{Ryu_2006} states that for special ``holographic states'', the entanglement of a subsystem $A$ is given by
\begin{equation}
S(A)= S_{\rm bulk}(\rho_a) + \frac{\mathscr{A}}{4G_N},
\end{equation}
which decomposes the entanglement of a boundary state in a subregion $A$ into two terms: the first representing the contribution from the bulk state, and the second from an area contribution of the Ryu-Takayanagi surface homologous to $A$. When viewing holography as an error-correcting code, the bulk state describing the matter fields is identified as the logical information, while the area term contains the shared entanglement of the code space makes up the underlying emergent geometry. These holographic properties can be exactly captured by stabilizer code constructions such as \cite{Pastawski:2015qua} and Fig.~\ref{fig:holo} where the minimal surface area that is captured by the stabilizer entanglement (edges in plane) is decoupled from the entanglement purely in the bulk state that contains all the magic. Our algorithm computes these entropies separately and then puts them back together. Notably, the RT formula that enables such a simplification in the entanglement computation also generalizes to any stabilizer codes even when $\mathscr{A}$ no longer admits a geometric interpretation \cite{Harlow2016TheRF,Pollack2021UnderstandingHE}. We will show that this decoupling allows us to efficiently compute the von Neumann and all R\'enyi entropies using the same algorithm. %Notably, the holographic error-correcting property extends beyond mere analogy. The stabilizer code itself is holographic. The associated ``complementary recovery'' property dictates a specific structure for the states within the code space, and enables two separate computations of entanglement in our algorithm.

A crucial assumption of our algorithm is the existence of a common stabilizer group $\mS$ within the stabilizer basis expansion; we also develop routines to extract $\mS$ when such information is not provided. For a superposition of stabilizers $|\Psi \rangle $, we design an algorithm to determine the maximal set of common stabilizers from the given $K$ stabilizer states. More excitingly, if a quantum computer prepares a state that is close to one with low stabilizer nullity, we can employ a learning algorithm\cite{Montanaro:2017oht,Grewal:2023hzn} to approximately identify the maximal common stabilizer set. This further allows us to apply our algorithm to learn the entanglement of low nullity states approximately. This approach presents a rare opportunity to measure the von Neumann entropy for a magical and highly entangled state in the experiments.

\section{Entropy in stabilizer codes}

We begin by introducing the holographic properties of a stabilizer code, the structure of the encoded states, and the RT formula to compute the entanglement.   

% Let $\mathcal P^n$ be the $n$-qubit Pauli group, then a stabilizer code\footnote{Most generally, stabilizer codes can also be defined by groups $S$ that are non-Abelian, but here we adapt the common naming convention where stabilizer codes refer to specifically Pauli stabilizer codes where $S$ is Abelian.} is identified by the subspace $\mathcal{C}$ of the $n$-qubit physical Hilbert space $\mathcal{H}_{\rm phys}$ uniquely identified by an Abelian stabilizer subgroup $S\subset \mathcal P^n$ such that $\mathcal{C}= \{|\psi\rangle\in \mathcal{H}_{\rm phys}:  \forall s\in S, s|\psi\rangle=|\psi\rangle\}$. 

Let $\mathcal P^n$ be the $n$-qubit Pauli group, and $\mS$ an Abelian subgroup hosting all the stabilizers. A stabilizer code\footnote{Most generally, stabilizer codes can also be defined by groups $\mS$ that are non-Abelian, but here we adapt the common naming convention where stabilizer codes refer to specifically Pauli stabilizer codes where $\mS$ is Abelian.}  is identified as a collection of states stabilized by the Pauli operators in $\mS$. These states form a code subspace  $\mathcal{C}$ of the $n$-qubit physical Hilbert space $\mathcal{H}_{\rm phys}$ with the definition $\mathcal{C}= \{|\psi\rangle\in \mathcal{H}_{\rm phys}:  \forall s\in \mS, s|\psi\rangle=|\psi\rangle\}$. It is known that an $[[n,k]]$ stabilizer codes that encodes $k$ logical qubits into $n$ physical qubits have $|\mS|=2^{n-k}$ and $|\mathcal C|=2^k$. 

Stabilizer codes are special in that they satisfy complementary recovery \cite{Pollack2021UnderstandingHE,Harlow2016TheRF}, a property of their logical operators that can be derived from e.g. the cleaning lemma \cite{Bravyi_2009,Kalachev_2022}. 
\begin{definition}[Complementary Recovery]
    Consider a bipartition of the $n$ physical qubits into complementary sets $A$ and $A^c$. Let $\mathcal{M}_A$ be the von Neumann algebra formed by logical operators that are supported on $A$, a code satisfies complementary recovery if all logical operators that commute with $\mathcal{M}_A$, i.e. its commutant $\mathcal{M}'_A$, are supported on $A^c$.
\end{definition}

%--- if we bipartition the $n$ physical qubits into complementary sets $A$ and $A^c$, then the logical operators supported on $A$ forms a logical subalgebra $\mathcal{M}_A$ and it commutes with the subalgebra supported on $A$ necessarily commute with that supported on $A^c$. 
%{\color{red}emphasize the definition of subalgebra, get rid of commutant}
It is instructive to first consider the special case of subsystem complementary recovery where the logical subalgebra $\mathcal{M}_A$ supported on $A$ is a factor, meaning that its center $Z_M=\mathcal{M}_A\cap \mathcal{M}_A'$ only contains elements proportional to the identity. For a stabilizer code encoding $k$ qubits, $\mathcal{M}_A$ being a factor means that it consists of logical operators acting on exactly $k_{a}\leq k$ logical qubits. Complementary recovery then states that $\mathcal{M}_{A^c}$ must contain the logical operators that act on the remaining $k_{a^c}=k-k_a$ logical qubits. For example, the perfect $[[5,1,3]]$ code falls under this category for any bipartition of the physical qubits because a subsystem $A$ either contains the support of both logical $X$ and the corresponding $Z$ operators or none at all, but never a single logical $X$ or a single logical $Z$ operator.

Harlow \cite{Harlow2016TheRF} showed that the above algebraic condition is equivalent to the following for state recovery --- for any codeword $|\bar{\psi}\rangle\in\mathcal{C}\subset \mathcal{H}_{\rm phys}$, there exist recovery unitaries $U_A$ and $U_{A^c}$, which only have support on $A$ and $A^c$ respectively, such that 
\begin{equation}
    U_A\otimes U_{A^c} |\bar{\psi}\rangle = |\psi\rangle_{A_1A_1^c}\otimes |\chi\rangle_{A_2A_2^c}
\end{equation}
where $A=A_1\cup A_2, A^c=A_1^c\cup A_2^c$. In other words, the encoded states of the $k_a$ qubits are recoverable on $A$ (the $A_1$ subsystem) while the remaining $k_{a^c}$ qubits are recoverable on $A^c$ (the $A_1^c$ subsystem). 

Defining $\rho_a=\Tr_{A_1^c}|\psi\rangle\langle \psi|$, $\chi=\Tr_{A_2^c}|\chi\rangle\langle\chi|$, then the entanglement entropy of $A$ is given by:
\begin{equation}\label{eq:RT}
    S(\rho_A) = S(\rho_a)+S(\chi)
\end{equation}
where the first entropy term on the right hand side is the von Neumann entropy of the logical subsystem over $k_a$ qubits and the second term captures the amount of entanglement resource that is present in the error correction code that enables non-trivial erasure correction. In the context of holographic error correction, $S(\rho_a)$ is precisely the bulk entropy while $S(\chi)$ captures the entropy associated with the minimal area surface in the RT formula. However, this formula holds for all stabilizer codes even if they do not admit geometric interpretations as in holography. 

Now recall the state of interest $|\Psi\rangle=\sum_{j=1}^Kc_j|\psi_j\rangle$ for which we hope to compute its von Neumann entropy where $|\psi_j\rangle$ are stabilizer states. If $\{|\psi_j\rangle, j=1,\dots K\}$ are orthogonal and stabilized by a common stabilizer group $\mS$ with $K\leq 2^k$, then $|\Psi\rangle\in \mathcal{C}$ is nothing but the codeword of a stabilizer code\footnote{Here a stabilizer group is common if and only if the elements of the group are identical, including the phase factors/signs.}. For convenience, one can further choose a basis for the logical operators of this code such that $|\psi_j\rangle$ are mapped to encoded computational basis states, but this is not required. Using the RT formula, we can now break down the entropy computation of $S(\rho_A)$ into an easy piece that is the RT entropy $S(\chi)$ which costs polynomial in $n$ and a hard piece that is the bulk entropy $S(\rho_a)$ which costs generally exponential in $k$. 

Since stabilizer codes are encoded using Clifford unitaries, any encoded stabilizer state, e.g. the computational basis state $|\bar{b}\rangle$, must be a stabilizer state in the physical Hilbert space. Choosing one such basis state for convenience and applying the RT formula above, one can first efficiently compute $S(\chi)=S(\Tr_{A^c}[|\bar{b}\rangle\langle\bar{b}|])$ using \cite{Fattal:2004frh} because the pre-encoded logical state $|b\rangle$ is a product pure state with zero bulk entropy contribution. Then to obtain $S(\rho_a)$ associated with the state $|\Psi\rangle$ of interest, we simply identify the logical basis states $|\bar{j}\rangle=|\psi_j\rangle$ such that $|\psi\rangle_{A_1A_1^c}=\sum_j c_j | j\rangle$ over $k$ logical qubits and $\rho_a = \Tr_{a^c}[|\psi\rangle\langle\psi|])$. For $K$ sufficiently small, $S(\rho_a)$ can then be obtained by brute force in the worst case. The hardness of the bulk entropy term is thus upper bounded by $\mathcal O(2^{\nu(|\Psi\rangle)})$  where
\begin{equation}
    \nu(\ket{\Psi})=n-\rank(\stab{\ket{\Psi}})
\end{equation}
is the stabilizer nullity\cite{beverland_lower_2020} of $|\Psi\rangle$ and  $\stab(\ket{\Psi})$ denotes the group of Pauli operators that stabilize $\ket\Psi$. For a stabilizer state, the nullity is zero, while for a state with no nontrivial Pauli stabilizers, the nullity is maximal and equal to $n$.  This quantity naturally captures the degree of non-stabilizerness or ``magic'' in a quantum state, and is known to be upper bounded by the number of non-Clifford gates in a circuit preparing the state \cite{Grewal:2023hzn}.

%In the context of holography in quantum gravity, this relation which 
%the RT formula applies when there exists a bipartition  $A\cup A^c$ of the physical degrees of freedom and local unitaries $U_A\otimes U_A^c$ such that 

%Using this stabilizer code perspective, we compute entanglement entropy by applying the Ryu-Takayanagi (RT) formula derived by Harlow \cite{Harlow2016TheRF}. This formula holds for all exact erasure-correcting codes \cite{PhysRevLett.94.180501,Kribs:2006doz,PhysRevA.75.064304}. 

More generally, however, the logical subalgebra $\mathcal{M}_A$ supported on a subregion $A$ may not be perfectly paired logical $X$ and $Z$s, namely not a factor. The simplest example of this is a two-qubit repetition code with $\mS=\langle ZZ\rangle, \bar{Z}=ZI, \bar{X}=XX$. Then for any one qubit subsystem, the logical subalgebra supported on $A$ is generated by $\bar{Z}$ only and is clearly not a factor but a direct sum of factors. 
Such stabilizer codes are still complementary in the most general subalgebra sense because $\M_A$ and $\M_{A^c}$ are both generated by $\bar{Z}$ and hence they trivially commute. 

%{\color{red}i need to think a bit more about how much detail to throw in here. It also depends on how the next section is written. Will add or remove stuff as needed later on.}
An analogous RT formula still applies to $\rho_A=\Tr_{A^c}\left(|\bar{\psi}\rangle\langle\bar{\psi}|\right)$:
\begin{equation}
    S(\rho_A)=S(\rho_a)+\mathscr{A}(\rho_A)
\end{equation}
%is defined as
%\begin{equation}\begin{aligned}
%    S(\rho_a)&:=S\lrp{\oplus_{\alpha}p_{\alpha}\rho_{a_{\alpha}}}\\
 %   &=-\sum_{\alpha}p_{\alpha}\log p_{\alpha}+\sum_{\alpha}p_{\alpha}S\lrp{\rho_{a_{\alpha}}}\\
   % \mathscr{A}(\rho_A)&=\sum_{\alpha}p_{\alpha}S(\chi_{\alpha}).
%\end{aligned} \end{equation}
where now $\rho_a$  takes a block diagonal form, one block for each factor, and is given by projecting the logical state $|\psi\rangle$ onto the logical subalgebra supported on $A$. % Both $\rho_a$ and the probabilities $p_{\alpha}$ depend on logical information $|\psi\rangle$ and the structure of the center $Z(\mathcal{M}_A)$. 
We provide a more detailed review of operator algebra quantum error correction and the more general form of Ryu-Takayanagi formula in Appendix~\ref{app:RT}.  

The ``area'' term $\mathscr{A}(\rho_A)$ is a generalization of $S(\chi)$ in Eq.~(\ref{eq:RT}) whose explicit form also depends on the block diagonal structure\footnote{We have absorbed all the constants into the definition of $\mathscr{A}$ so it becomes a dimensionless quantity.}.  However, it is easy to compute it in stabilizer codes:
\begin{theorem}
\label{th:stab_code_area_term}
    The area contribution $\mathscr{A}(\rho_A)$ for any stabilizer code is independent of the encoded logical state \cite{Cao2023}.
\end{theorem} 
Therefore, a similar but more general procedure to compute the bulk entropy and ``area'' still applies where the former can be obtained by brute force while the latter can be efficiently computed by setting the logical state to a stabilizer reference state and computing the total entanglement entropy of $A$. 

In particular, because the area contribution is entirely stabilizer and bulk contribution captures all the non-stabilizerness which we assume to be low, our algorithm also returns the full entanglement spectrum, i.e., the singular values of $\rho_A$, and the R\'enyi entropies efficiently. We explain how this can be done in detail in Appendix~\ref{app:Renyi}. This is a notable advantage of our algorithm compared to the earlier algorithms\cite{gu_doped_2024,gu_magic-induced_2025,aziz_classical_2025} of computing entanglement on low magic states, which only focus on R\'enyi entropy with integer index greater than or equal to $2$.

\section{Efficient algorithm for entanglement entropy}

Since our algorithm is based on the RT formula \cite{Harlow2016TheRF,Pollack2021UnderstandingHE,Cao2023}, where $S(\rho_a)$ is the entropy of the logical state and $\mathscr{A}(\rho_A)$ is the area term, we detail the procedures to compute each of them below. 
\begin{algorithm}[H]
\caption{Compute $S(\rho)$ for a State with Nullity $\nu$}
\label{alg:combined}
\begin{algorithmic}[1]
\Function{compute\_logical\_entropy}{$\mS, A, |\psi\rangle$}
    \State \textbf{Input:} Generators of the common stabilizer group $\mS$, subsystem $A$, state $|\psi\rangle$.
    \State \textbf{Output:} The logical entropy $S(\rho_a)$.
    \State $\{\bar{X}_i, \bar{Z}_i\}_{i=1}^{2\nu} \gets $ \text{$2\nu$ logical operators.}
    \State $\{\bar{P}_j\}_{j=1}^{k_a} \gets$ \text{$k_a$ logical operators in $A$}.
    \State $\rho_{a} \gets \text{zeros}(2^{k_a}, 2^{k_a})$
    \For{each logical Pauli operator $\bar{P}$ in A}
        \State $\rho_{a} \gets \rho_{a} + \frac{1}{2^{k_a}} \cdot \langle \psi | \bar{P}| \psi \rangle \cdot P$
    \EndFor
    \State $S(\rho_a)\gets -\text{Tr}(\rho_{a} \ln \rho_{a})$
\EndFunction

\Function{compute\_entanglement}{$|\Psi\rangle, A$}
    \State \textbf{Input:} State $|\Psi\rangle = \sum_j c_j |\psi_j\rangle$, Subsystem $A$.
    \State \textbf{Output:} von Neumann Entropy $S(\rho_A)$.

    \State $\mS \gets $\text{Find common stabilizers (dimension = $n-\nu$).}
    \State $S(\rho_a) = $ \text{compute\_logical\_entropy}{$(\mS, A, |\Psi\rangle)$}
    \State $|\phi \rangle \gets$ \text{reference state stabilized by $\mS\cup \{\bar{Z}_i\}_{i=1}^{\nu}$}
    \State $S_A( |\phi \rangle ) \gets$ entanglement of $A$ for $|\phi \rangle $.   
    \State $S(\phi_a, |\phi \rangle) = $ \text{compute\_logical\_entropy}{$(\mS, A, |\phi\rangle)$}
    \State $\mathscr{A} = S_A( |\phi \rangle ) - S(\phi_a, |\phi \rangle)$
    \State $S(\rho_A) = S(\rho_a) + \mathscr{A}$
\EndFunction
\end{algorithmic}
\end{algorithm}

{\bf Computation of $S(\rho_a)$}: This is the non-stabilizer part of the state coming from the $2^{\nu}$ dimensional logical space. We compute it via a full reconstruction of the logical state $\rho_a$: 
\begin{itemize}[itemindent=2.8em, left=2.8em]
\item[\bf Step 1] \underline{Logical operators in the subregion $A$}. From the $n - \nu$ common generators in $\mS$, we construct a complete set of $2\nu$ logical operators $\{ \bar{X}_i, \bar{Z}_i \}_{i=1}^{\nu}$, which commute with $\mS$ but are not elements of it. Through gauge transformation by multiplications by stabilizers, we can find a generating set of logical operators that are completely in $A$. They form von Neumman algebra $\mathcal{M}_A$.  
\item[\bf Step 2] \underline{Logical state $\rho_a$}: Through complementary recovery, the $2^{k_a} \times 2^{k_a}$ logical density matrix elements can be reproduced in the code subspace: $\tr( \rho_a P_i) = \langle \Psi | \bar{P}_i | \Psi \rangle$, where $\bar{P}_i$ is the physical representation of the logical operators in $P_i$. The expectation values can be computed efficiently by the Gottesman-Knill theorem \cite{Gottesman1998TheHR, aaronson_improved_2004} as long as $\nu(|\Psi\rangle)$ is at most $\mathcal{O}(\log n)$. 
\item[\bf Step 3] \underline{Algebraic entropy}: Since the dimension of $\rho_a$ is $2^{k_a} \times 2^{k_a}$, a brute force computation of $S(\rho_a)$ through diagonalization has complexity $\mathcal O(2^{3 k_a} ) \lesssim \mathcal O(8^{\nu})$.
\end{itemize}

The center of the algebra can bring in a block diagonal structure of $\rho_a$. This can further reduce the complexity of Step 3, but not in the worst case. A detailed analysis is given in future work when we consider low entangled logical states.

{\bf Computation of the area term}: 
Next, we invoke Theorem~\ref {th:stab_code_area_term}: since the area term is independent of the logical state, we use a reference stabilizer state to compute it. 
\begin{itemize}[itemindent=2.8em, left=2.8em]
\item[\bf Step 4] \underline{Reference state}: Choose any stabilizer state $|\phi \rangle$ in the code subspace. One choice is the state defined by the stabilizers with generators in $S$ and $\{ \bar{Z}_i\}_{i = 1}^{\nu}$. 
\item[\bf Step 5] \underline{Area term}: Compute the $S(\phi_A)$ through the standard Gottesman-Knill algorithm. Invoke the algorithm above to compute the algebraic entropy $S(\rho_a( \phi ) )$. Subtract to obtain the area law term $\mathscr{A} = S(\phi_A) - S(\rho_a( \phi ) )$.
\end{itemize}

As an illustration,  consider the following example state as a superposition of two stabilizer states: 
\begin{equation}\begin{aligned}
    \ket{\psi}=&\frac{c_1}{\sqrt{2}}\left(\ket{0000}+\ket{1111}\right)+\frac{c_2}{\sqrt{2}}\left(\ket{0101}+\ket{1010}\right),
\end{aligned} \end{equation}
where the common stabilizer group $\mS$ is generated by $XXXX$, $IZIZ$ and $ZIZI$, corresponding to the $[[4,1,2]]$ stabilizer code. Without loss of generality, we take the first two qubits as the subregion $A$.

% Following the algorithm outlined in the main text, we first compute the logical entropy $S(\psi_a)$. This requires applying the Gaussian elimination decoder to determine the algebra that can be recovered from the subregion $A$.

We begin by writing the stabilizer generators in their binary symplectic representation. For clarity, the corresponding Pauli operators are listed in the top row:
\begin{equation}
    H_{S}=\left(
    \begin{array}{cccc|cccc|c}
        &&Z&&  &&X&&\\
        0 & 0 & 0 & 0 & 1& 1 &1 &1 &S_1\\
        0 &1 &0 &1& 0& 0&0 &0&S_2\\
        1&0&1&0&0&0&0&0&S_3
    \end{array}
\right).
\end{equation}
In Step 1, we apply Gaussian elimination to compute the kernel of $H_S$, thereby obtaining the binary symplectic vectors corresponding to the logical operators. Among these, we identify a logical operator $\bar{Z}$ that has support only on the subregion $A$:
% \begin{equation}
% H_L=\left(
%     \begin{array}{cccc|cccc|c}
%     &&Z& & &&X&\\
%      0 & 1 & 1  & 0 & 0& 0 &0 &0 & \bar{Z}\\
%      0 & 0 & 0  & 0 & 1& 0 &1 &0 & \bar{X}
%     \end{array}
% \right).
% \end{equation}

\begin{equation}\begin{aligned}
    H_{LA}=\left(
    \begin{array}{cccc|cccc|c}
    &&Z& & &&X&\\
     1 & 1 & 0  & 0 & 0& 0 &0 &0 & \bar{Z}
    \end{array}
\right).
\end{aligned} \end{equation}
In Step 2, the reduced logical state $\rho_a$ is obtained by reconstruction as
\begin{equation}\begin{aligned}
    \psi_a=\frac{1}{2}\left(I+\bra{\psi}ZZII\ket{\psi} Z\right)
    =\frac{1}{2}\left(I+(|c_1|^2-|c_2|^2)Z\right). 
\end{aligned} \end{equation}
% Note that since $\ket{\psi}$ is a superposition of at most $2^k$ stabilizer states, the expectation value can always be obtained with complexity upper bounded by $O(e^k\poly(n))$.
In step 3, we compute the algebraic entropy as  
\begin{equation}\begin{aligned}\label{eq:lentropy}
    S(\psi_a)=&-\left(\frac{1}{2}+\frac{|c_1|^2-|c_2|^2}{2}\right)\ln\left(\frac{1}{2}+\frac{|c_1|^2-|c_2|^2}{2}\right)\\
    &-\left(\frac{1}{2}-\frac{|c_1|^2-|c_2|^2}{2}\right)\ln\left(\frac{1}{2}-\frac{|c_1|^2-|c_2|^2}{2}\right).
\end{aligned} \end{equation}
In step 4, we choose a reference state $|\phi \rangle$ by setting $c_2 = 0$ in $|\psi \rangle$,
\begin{equation}\begin{aligned}
    \ket{\phi}=\frac{1}{\sqrt{2}}\left(\ket{0000}+\ket{1111}\right),
\end{aligned} \end{equation}
whose stabilizer group are generated by $S \cup \bar{Z}$. 

In Step 5, we first compute the entanglement entropy of the first two qubits, obtaining $S_A(\phi)= \ln 2$. From Eq.~\eqref{eq:lentropy}, we find that when $c_2 = 0$, the logical entropy of $\ket{\phi}$ reduces to $S(\phi_a)=0$. It follows that the area term is $\mathscr{A}(\rho)=S(\phi_A)-S(\phi_a)=\ln 2$. Therefore, the entanglement entropy of the state $\ket{\psi}$ is
\begin{equation}\begin{aligned}
    S(\psi_A)=S(\psi_a)+\ln 2.
\end{aligned} \end{equation}

\section{Extract the maximal stabilizer set}

Our algorithm requires two subroutines: a method to extract the maximal common stabilizer group $\text{STAB}(|\Psi \rangle)$, and the evaluation of the Pauli expectation value for $|\Psi \rangle$. It is well-known that the latter can be done efficiently using the stabilizer formalism as long as $|\Psi\rangle$ is written as a superposition of stabilizer states with small enough $K$. We now detail the first subroutine for (1) a given superposition and (2) quantum states promised to be close to the low nullity manifold. 

A Pauli operator $P$ is a stabilizer if the expectation value $\langle \Psi | P |\Psi \rangle = \pm 1$. Thus the stabilizers can be discovered from the Pauli spectrum. The following method produces the Pauli spectrum for a superposition of stabilizers. To illustrate, consider the case of two non-orthogonal states, $|\Psi \rangle = a | s_1 \rangle + b | s_2 \rangle$ where $|s_1\rangle$ and $|s_2\rangle$ are stabilizer states with stabilizer groups $\mathcal{S}_1$ and $\mathcal{S}_2$, respectively. Define projectors $\rho_j = |s_j\rangle \langle s_j |$ for $j = 1, 2$. Since both are density matrices of stabilizers, we have $\rho_j = \frac{1}{2^n}\sum_{g\in \mathcal{S}_j} g_j$. In other words, they are an equal-weight superposition of the stabilizers. The density matrix of the superposition can be written as
\begin{equation}
\begin{aligned}
     |\psi \rangle \langle \psi | &= |a|^2 |s_1 \rangle \langle s_1| 
     + |b|^2 |s_2 \rangle \langle s_2| \\
     &+  \left(\frac{a b^*}{\langle s_1| s_2 \rangle} |s_1 \rangle \langle s_1 | s_2 \rangle  \langle s_2|  + \text{h.c.} \right)   \\
     &= |a|^2 \rho_1 + |b|^2 \rho_2 + \left( \frac{a b^*}{\langle s_1| s_2 \rangle} \rho_1 \rho_2 + \text{h.c.} \right). 
\end{aligned}    
\end{equation}
Expand $\rho_j$ by the stabilizers, we see that each of the four terms above gives an equal weight superposition of Pauli matrices. These Pauli matrices are in the form $g_1 g_2$ where $g_1 \in \mS_1$, and $g_2\in \mS_2$. Due to their degeneracies of weights, it is sufficient to compute the coefficient of a single representative to determine the expectation values of all operators in that class, as well as to identify which of them are stabilizers. In this way, we can extract the Pauli spectrum and read out the maximal stabilizer group. See Appendix~\ref{app:superposition} for details. 

In the case when the two states are orthogonal, then there is a common stabilizer $g$ such that $\bra{s_1}g|s_1\rangle=-\bra{s_2}g|s_2\rangle=\pm1$. In other words, the states share the same stabilizer elements but with the opposite eigenvalues. We can replace $\langle s_1 | s_2 \rangle$ in the denominator by $\langle s_1 | E | s_2 \rangle$ where $E$ is a suitably chosen error operator that flips the eigenvalues of all such stabilizers, ensuring that $\langle s_1 | E | s_2 \rangle \neq 0$. The Pauli spectrum can again be read out from the density matrix expression. Details of the more general case, including a superposition of $K$ states are deferred to Appendix~\ref{app:generalcase}.

Next, we consider that the state $|\Psi \rangle$ is prepared on a quantum device, and multiple copies of $|\Psi \rangle$ are accessible. In order to learn the entanglement spectrum using our algorithm, it demands a quantum algorithm to extract the maximal stabilizer group $\mS:=\stab(\ket{\Psi})$. We adapt an existing learning algorithm \cite{Montanaro:2017oht,Grewal:2023hzn} to extract $\mS$.

The basic idea of learning $\mS$ is that quantum measurements can automatically expose the Pauli operators that commute with $\mS$, namely, those in the normalizer $N(\mS)$. Pauli operators can be classified into three categories: the stabilizers in $\mS$, the logical operators in $N(\mS)/\mS$, and the error operators in $\mathcal P^n / N(\mS)$, with $\mathcal P^n$ denoting the n-qubits Pauli group.  An error operator $E$ must anticommute with at least one stabilizer $g \in S$. Hence the expectation value $\langle \Psi | E | \Psi \rangle$ is $\frac{1}{2} \langle \Psi | \{ E, g\}  | \Psi \rangle  = 0$. Therefore, only Pauli operators in $N(\mS)$ have non-zero expectation values. 

In the Bell difference sampling\cite{Montanaro:2017oht}, the expectation values of Pauli operators are transformed into probabilities of projective measurements of Bell states. The wavefunction collapse reveals those operators with non-zero expectation values, namely those in $N(\mS)$. There is a classical algorithm of complexity $\mathcal O(n^3)$ to revert the normalizer $N(\mS)$ back to $\mS$. Our algorithm can take the quantum measurement of $\mS$ as a subroutine, followed by logical operator measurements to reconstruct $\rho_a$. With these procedures, it can effectively learn the von Neumann entropy and entanglement spectrum of $|\Psi \rangle$. 

The above assumes a perfect extraction of $\mS$ with sufficient samples. In practice, finite sampling may lead to an overestimated stabilizer group $\hat{\mS}$ --- a superset of the true $\mS$—and hence to an approximate logical state $\hat{\rho}_a$. To estimate the error, we use the reconstructed state $|\hat{\Psi} \rangle$ in the learning algorithm\footnote{The algorithm does not stop at extracting $\hat{\mS}$; it constructs a full classical description of an approximate state $|\hat{\Psi }\rangle = \hat{C} | \text{logical} \rangle |\text{syndrome} \rangle$} of Ref.~\cite{Grewal:2023hzn}, that is $\epsilon$ close to the true state $|\Psi \rangle$ in trace distance if we can afford $\text{poly}(n)\mathcal O(1/\epsilon)$ samples. The intermediate data of our algorithm, the maximal stabilizer group $\hat{\mS}$ and the logical state $\hat\rho_a$, can be viewed as being extracted from the fully reconstructed state $|\hat{\Psi }\rangle$. By applying Fannes-Audenaert continuity\cite{Fannes1973,Audenaert_2007} bound of the entanglement entropy, the estimated entropy $S_A ( \hat{\Psi} )$ is $\mathcal O(n\epsilon)$ close to $S_A( \Psi )$. Reducing the error of entanglement to $\mathcal O(\epsilon)$ still requires $\text{poly}(n)\mathcal O(1/\epsilon)$ samples.  See Appendix~\ref{app:quantumlearn} for a detailed review and discussion of the quantum algorithm. 

\section{Applications} 

Our algorithm excels in the complementary regime accessible by the matrix product state \cite{white_numerical_1993,white_density-matrix_1993,schollwoeck_density-matrix_2011,schuch_entropy_2008}: while the latter is limited by at most logarithmic entanglement, our algorithm can deal with states of potentially high entanglement but is limited by at most logarithmic magic (characterized by nullity). A useful byproduct of the structure analysis in our algorithm produces all the R\'enyi entanglement entropy and the entanglement spectrum. 

An immediate application is to compute the entanglement of states generated by T-doped or magic-augmented Clifford circuits\cite{bravyi_simulation_2019}. When the amount of doping is sparse (e.g. bounded by $\mathcal O(\ln n)$ ), the resulting states not only have low stabilizer rank (small $K$) but also small nullity (small $\nu$) --- just the kind of states our algorithm can deal with. See Appendix~\ref{app:application} for detailed discussion.

There are many scenarios in quantum information and quantum many-body physics where this is useful. For example, the preparation of approximate unitary $k$ design (additive error) is shown to be possible with only $\mathcal O(k^4)$ magic gates \cite{Zhang:2025dhg,Haferkamp:2020qel}, which is independent of the system size $n$.

Although the worst case complexity of our algorithm is quantified by the stabilizer nullity, it remains efficient for many classes of states where both entanglement and magic are high, e.g. of $\mathcal{O}(n)$. For  example, a class of such states are codewords of holographic stabilizer codes (or higher dimensional multi-scale entanglement renormalization ansatz \cite{2dmera} with isometries and unitaries generated by Clifford unitaries) with $D$ bulk spatial dimensions where all the bulk/logical qubits are tensor product of non-stabilizer states. Such states have volume law magic and $\mathcal{O}(n^{\frac{D-1}{D}})$ entanglement and yet its entanglement entropy of  subregions are efficiently computable using our algorithm. More generally, the algorithm applies to a wider class of states generated by magic augmented Cliffords where magic gates are only applied at the beginning or the end of the Clifford. Interestingly, the algorithmic efficiency is not set by the total magic or entanglement, where both can scale as $\mathcal{O}(n)$ --- instead, it is lower bounded by the size of the ``non-canonical'' logical subalgebra in the stabilizer code defined by the Clifford circuit. We describe the details of the adapted algorithm and its limitations in Appendix~\ref{app:application}. 

Another key application is the research of the measurement-induced phase transition\cite{li_quantum_2018,skinner_measurement-induced_2019,chan_unitary-projective_2019,choi_quantum_2020,jian_measurement-induced_2020,gullans_entanglement_2019,ippoliti_entanglement_2020,fisher_random_2023}, where entanglement is the order parameter to characterize the transition. Large-scale numerics of MIPT has been largely limited to Clifford circuits\cite{li_quantum_2018,gullans_entanglement_2019,li_conformal_2020,li_measurement-driven_2019}, where the absence of magic leads to non-generic behaviors such as a flat entanglement spectrum. Our algorithms enable the simulation of circuits with both Clifford and non-Clifford gates interspersed with Pauli measurements\cite{aziz_classical_2025}, where these measurement tends to increase the stabilizers and suppress the nullity. This opens the door to investigating the interplay between entanglement and magic dynamics across the phase transition. 

Our method is also applicable to entanglement properties of the resource states for quantum computing. These states are often created by injecting magic into highly entangled stabilizer states\cite{raussendorf_one-way_2001} (e.g. 2D topological code\cite{fowler_surface_2012}). Our algorithm can operate in any dimension and provides full access to the entanglement spectrum. This allows for a direct numerical test of proposals like the Li-Haldane conjecture\cite{li_entanglement_2008}, which relates the entanglement spectrum to the edge excitation modes. 

More broadly, the general learnability of entanglement in quantum states has a variety of applications in quantum experiments, where entanglement has been used as a key diagnostic for quantum phases and emergent geometry in quantum gravity. Even though near-stabilizer quantum states have been shown to be efficiently learnable, there has been no definitive work demonstrating that von Neumann entropy, and in fact, the entire entanglement spectrum, can also be efficiently extracted as a result. Our work now confirms that this is indeed the case.

% Finally, our algorithm is one of the first kinds to measure the von Neumann entropy of quantum states prepared in an experiment, given that the state is close to a low nullity state. In a quantum computer operates on a stabilizer code itself, the physical states are assumed in the code space, thus satisfying the input assumptions of our algorithm. 

\section{Conclusion and Discussion}

In this work, we introduce an efficient classical algorithm to compute the entanglement for many-body states that permit volume-law entanglement but are magic-scarce. It represents the first step towards numerically investigating entanglement for magic states and provides a useful tool for studying quantum many-body systems. 

There are also many potential avenues to improve  our algorithmic efficiency. 

Although the main bottleneck of the algorithm is stabilizer nullity, only the  non-local nullity contribution $\nu^{NL}(\psi)=\min_{U_A,U_{\bar A}}\left( U_A\otimes U_{\bar A}\ket{\psi} \right)$ is unavoidable when it comes to entanglement computations. This is because entanglement entropies are invariant under local unitary (LU) transformations, hence the removal of local magic can produce a LU-equivalent state $|\Psi'\rangle$ with smaller stabilizer nullity but identical entanglement, leading to more efficient entanglement calculations. %Although the complexity bottleneck now has been reduced to non-local nullity $\nu^{NL}(\psi)=\min_{U_A,U_{\bar A}}\left( U_A\otimes U_{\bar A}\ket{\psi} \right)$, 
Nevertheless, how local magic should be removed in practice remains an open challenge which we shall leave for future work.

% in $|\Psi\rangle$  Therefore it might be useful to define the non-local stabilizer nullity as 

%\begin{definition}
%    Non-local stabilizer nullity is defined as the minimal smoothed stabilizer nullity of the state deformed by LU transformation. 
 %   \begin{equation}\begin{aligned}
 %       \nu^{NL}(\psi)=\min_{U_A,U_{\bar A}}\left( U_A\otimes U_{\bar A}\ket{\psi} \right).
%    \end{aligned} \end{equation}
%\end{definition}

The measure of magic also matters --- the complexity of stabilizer simulations is typically limited by the stabilizer rank, which lower bounds nullity. However, our algorithm's efficiency relies on the states of interest having low stabilizer nullity instead of stabilizer rank. For example, it can take exponential time for even a simple case like $a|0 \rangle^{\otimes n} + b | + \rangle ^{\otimes n}$, which has a low stabilizer rank but maximal nullity. Thus, determining whether efficient extensions to states with low stabilizer rank constitutes another important direction to explore. 

To further boost the algorithm's efficiency and thus widen its applicability, we can exploit additional structures present in the code or the encoded state. For example, it is known that (generalized) concatenated codes with linear rate $k/n$ allows one to construct $U_A, U_{A^c}$ using a sequence of smaller recovery unitaries where at each step only a small subset of logical qubits are recovered at a time. If every such subset of qubits are only weakly entangled with other logical qubits, then one can apply the RT-like decomposition iteratively where multiple bulk entropy terms are obtained in sequence, thus circumventing the exponential cost in computing the entropy of the bulk state living in the $2^k$ dimensional Hilbert space all at once. 

Relatedly, one can consider logical states that permit efficient classical representation, such as the matrix product state. Such representation removes the small $\nu$ constraints while still maintaining the efficiency of our algorithm. This structure is related to the recently proposed Clifford-Augmented MPS (CAMPS) algorithm\cite{qian_augmenting_2024-1,paviglianiti_estimating_2024,oliviero_unscrambling_2024,niroula_phase_2024,liu_classical_2024,huang_non-stabilizerness_2024,huang_clifford_2024,fux_disentangling_2025,frau_stabilizer_2024}, which can efficiently compute the correlation function, but not entanglement. 

Finally, the main drive for computational speedup comes from the stabilizer formalism, where properties of states are encoded as much simpler group structures and transformations. Nevertheless, the Pauli stabilizer formalism is only a special instance of the stabilizer formalism at large, where non-Abelian stabilizer groups and non-Pauli stabilizer states can and have been constructed \cite{Webster_2022,Ni_2015}. For such states, a check matrix extension can also be examined where some non-Clifford processes can be simulated efficiently \cite{Webster_2022,Shen:2023xmh}. One can ask whether the Pauli stabilizer technique for computing the entanglement entropy similar to \cite{Fattal:2004frh} can be analogously extended to other non-Pauli stabilizer states such as the XS or XP stabilizer states which are ``magical''. If that is the case, then generalizations of our technique to non-Pauli stabilizer codes that satisfy complementary recovery can similarly be used to compute entanglement also in magic-rich states.

\section{Acknowledgment}
We thank Cathy Li for helpful comments and discussions. T.Z. acknowledges Xiao Chen for presenting a second R\'enyi entropy formula in the same spirit as Eq.(3) of Ref.~\cite{gu_magic-induced_2025} in the initial stage of the project.

\appendix
\section{Ryu-Takayanagi Formula in Operator Algebra Quantum Error Correction}
\label{app:RT}
%\subsection{Review of Ryu-Takayanagi formula in QECC}

In the main text, we calculate the entanglement entropy of a codeword state using the Ryu-Takayanagi (RT) formula. Here, we review the general requirement, the complementary recovery of logical operator algebras in bipartitions, that is necessary for a quantum error-correcting code (QECC) to satisfy the (two-sided) RT formula. 

Consider a QECC with code subspace $\mathcal{C}\subset \mathcal{H}_{\text{phys}}$. Let $\mathcal{H}_{L}$ denotes the logical Hilbert space, which is isomorphic to $\mathcal{C}$. Define an encoding isometry $V: \mathcal{H}_L\rightarrow \mathcal{H}_{\text{phys}}$, such that the image of $V$ is exactly $\mathcal{C}$. 

Let $A$  be a subsystem of the physical qubits, inducing a factorization of the physical Hilbert space as $\mH_{\text{phys}}=\mH_{A}\otimes \mH_{A^c}$, where $A^c$ is the complement of $A$. The algebra of operators supported on $A$,  is denoted by $\mathcal{L}(\mathcal{H}_{A})$. Projecting these operators onto the logical subspace via the encoding isometry $V$, one defines the logical operator algebra as,   
\begin{equation}
    \mathcal{M}_A:=V^{\dagger}\mathcal{L}(\mathcal{H}_{A})\otimes I_{ A^c}V \subset \mathcal{L}(\mathcal{H}_L) 
\end{equation}

If $\mathcal{M}_A$ forms a \textit{von Neumann algebra} -- that is, it is closed under multiplication, scalar multiplication,  addition, and complex conjugation -- then the triple  $(V,A,\mathcal{M}_A)$ satisfies the property known as \textit{complementary recovery}.  Specifically, this means that the commutant of $\mathcal{M}_A$, denoted as $\mathcal{M}'_A$,  can be recovered on the complementary subsystem $\bar A$.  (For the proof, see \cite{Pollack2021UnderstandingHE}). 

Given that $\mathcal{M}_A$ is a von Neumann algebra acting on the logical Hilbert space $\mathcal{H}_L$, the \textit{classification theorem} guarantees that there exists a block decomposition of the form 
\begin{equation}
    \mathcal{H}_L=\bigoplus_{\alpha}\mathcal{H}_{a_\alpha}\otimes \mathcal{H}_{a_\alpha^c},
\end{equation}
such that the algebra $\mathcal{M}_A$ and its commutant $\mathcal{M}_A'$ take the standard forms
\begin{equation}
\begin{aligned}
    \mathcal{M}_A&=\bigoplus_{\alpha} \mathcal{L}(\mathcal{H}_{a_{\alpha}})\otimes I_{ a_\alpha^c}\\
    \mathcal{M}'_A&=\bigoplus_{\alpha}I_{a_{\alpha}}\otimes \mathcal{L}(\mathcal{H}_{a_{\alpha}^c}).
    \end{aligned}
\end{equation}

At the operator level, any element in $\mathcal{M}_A$ of the form $\oplus_{\alpha}O_{a_\alpha}\otimes I_{ a_\alpha^c}$, or in $\mathcal{M}'_A$ of the form $\oplus_{\alpha}I_{ a_\alpha}\otimes O_{a_\alpha^c}  $ , admits a reconstruction in the physical Hilbert space. Specifically, there exist operators $W_A\otimes I_{ A^c}$ , or $I_A\otimes W_{ A^c}$ acting on $\mathcal{H}_{\text{phys}}$ such that
\begin{equation}
\begin{aligned}
    V^{\dagger}W_A\otimes I_{ A^c}V=\bigoplus_{\alpha}O_{a_\alpha}\otimes I_{a_\alpha^c}\\
    V^{\dagger}I_A\otimes W_{A^c}V=\bigoplus_{\alpha}I_{ a_\alpha}\otimes O_{a_\alpha^c}.
    \end{aligned}
\end{equation}

At the state level, complementary recovery ensures that logical information can be recovered through local unitaries supported on $A$ and $ A^c$. Let  $\{\ket{\alpha,i}_{a_{\alpha}}\otimes\ket{\alpha,j}_{a_{\alpha}^c}\}$ be a basis spanning each $\alpha$-block of the logical Hilbert space ($\mathcal{H}_{a_\alpha}\otimes \mathcal{H}_{a_\alpha^c}$). The corresponding encoded state is 
\begin{equation}
    \ket{\widetilde{\alpha,ij}}=V\ket{\alpha,i}_{a_{\alpha}}\ket{\alpha,j}_{a_{\alpha^c}}.
\end{equation}

There exists a pair of local unitaries $U_A$ and $U_{A^c}$,  such that for all codeword states,
\begin{equation}\begin{aligned}
    U_A\otimes U_{ A^c}\ket{\widetilde{\alpha,ij}}_{AA^c }=\ket{\alpha,i}_{{A_1}_{\alpha}}\ket{\alpha,j}_{ {A_1}^c_{\alpha}}\ket{\chi_{\alpha}}_{{A_2}_{\alpha} {A_2}^c_{\alpha}},
\end{aligned} \end{equation}
where the physical Hilbert space decomposes as
\begin{equation}
\begin{aligned}
    \mathcal{H}_{A}=\lrp{\bigoplus_{\alpha}\mathcal{H}_{A_1^{\alpha}}\otimes \mathcal{H}_{A_2^{\alpha}}}\oplus \mathcal{H}_{A_3},
\end{aligned} 
\end{equation}
with $\mathcal{H}_{A_1^{\alpha}}\cong \mathcal{H}_{a_{\alpha}}$, and similarly for the complementary subsystem.

A consequence of this factorization structure is that for any state $\ket{\psi}\in \mathcal{C}$,  the reduced density matrix can be written as 
\begin{equation}\begin{aligned}
    \rho_{A}=&\Tr_{\bar A}(\ket{\psi}\bra{\psi})\\
    =&U_A\lrp{\oplus_{\alpha}p_{\alpha}\rho_{a^{\alpha}}\otimes \chi_{\alpha}}U_{A}^{\dagger}
\end{aligned} \end{equation}
where $p_{\alpha}\rho_{a_{\alpha}}=\Tr_{{ a_{\alpha}^c}}\lrp{\Pi_{\alpha}V^{\dagger}\ket{\psi}\bra{\psi}V\Pi_{\alpha}}$, and $\chi_{\alpha}=\Tr_{ {A_2}^c_{\alpha}}\lrp{\ket{\chi_{\alpha}}\bra{\chi_{\alpha}}}$. Then the entropy of $\rho_A$ has the following decomposition:
\begin{equation}\begin{aligned}
    S(\rho_A)&=S(\rho_a)+\mathscr{A}(\rho)
\end{aligned} \end{equation}
where $S(\rho_a)$ and $\mathscr{A}(\rho_A)$ are defined as
\begin{equation}\begin{aligned}
    S(\rho_a)&:=S\lrp{\oplus_{\alpha}p_{\alpha}\rho_{a_{\alpha}}}\\
    &=-\sum_{\alpha}p_{\alpha}\log p_{\alpha}+\sum_{\alpha}p_{\alpha}S\lrp{\rho_{a_{\alpha}}}\\
    \mathscr{A}(\rho_A)&=\sum_{\alpha}p_{\alpha}S(\chi_{\alpha})
    % \Tr(\rho \mathcal{L}_A)\\
    % \mathcal{L}_A&=\oplus_{\alpha} p_{\alpha}S\lrp{\chi_{\alpha}}I_{a_{\alpha}\bar a_{\alpha}}
\end{aligned} \end{equation}

This is derived in \cite{Harlow2016TheRF} and is referred to as the Ryu-Takayanagi (RT) formula in QECC. This is analogous to the RT formula in holographic theory \cite{PhysRevLett.96.181602}, where $S(\rho_a)$ represents the entropy of matter fields while $\mathscr{A}(\rho_A)$ plays the role of minimal surface area of the spacetime. 

As a special class of QECC, stabilizer code plays an important role and is proved to satisfy the  complementary recovery \cite{Pollack2021UnderstandingHE}. It was further proved in \cite{Cao2023} that the term $\mathscr{A}(\rho)$ for stabilizer code is a state independent factor, and therefore being the same for all the state in the code subspace. 
Often literature also differentiate the type of complementary recovery based on the structure of the subalgebra $\mathcal M_A$. Suppose complementary recovery applies when $\mathcal{M}_A$ is a factor, then the code is said to satisfy \textit{subsystem complementary recovery}. The most general form we discussed above is referred to as \textit{subalgebra complementary recovery}. 

\section{Extracting the entanglement spectrum}
\label{app:Renyi}
The algorithm also extends to R\'enyi entropies, since an RT-like formula applies in this case as well:
\begin{equation}\begin{aligned}
    S_{n}(\rho_A)=S_n(\rho_a)+\mathscr{A}
\end{aligned} \end{equation}

where $S_n$ denotes the nth R\'enyi entropy while $\mathscr{A}$ is a rescaled area-like term which turns out to be independent of $n$. 

To see this, recall from Ref.~\cite{Harlow2016TheRF} that the reduced state on $A$ admits the block-diagonal decomposition:
\begin{equation}\begin{aligned}
U_A\rho_AU_A^{\dagger}=\oplus_{\alpha} p_{\alpha}\rho_{a^{\alpha}}\otimes \chi_{\alpha}.    
\end{aligned} \end{equation}
Given this decomposition, the R\'enyi entropy of it can be computed as
\begin{equation}\begin{aligned}
    S_{n}(\rho_A)=&\frac{1}{1-n}\log\Tr(\rho_A^n)\\
    =&\frac{1}{1-n} \log\lrp{\sum_{\alpha}p_{\alpha}^n\Tr(\rho_{a^{\alpha}}^n)\Tr(\chi_{\alpha}^n)}
\end{aligned} \end{equation}

It was shown in Ref.~\cite{Cao2023} that for any pair of states $\chi_{\alpha}$ and $\chi_{\beta}$, there exists a Pauli string operator $P$ such that 
\begin{equation}\begin{aligned}
    P\chi_{\alpha}P^{\dagger}=\chi_{\beta}.
\end{aligned} \end{equation}

Thus, the $\chi_\alpha$ are iso-spectral, allowing us to factor out the common contribution $\Tr(\chi_{\alpha}^n)$ and obtain:
\begin{equation}\begin{aligned}
    S_n(\rho_A)=\frac1{1-n}\log\lrp{\sum_{\alpha}p_{\alpha}^n\Tr(\rho_{a^{\alpha}}^n)}+\frac{1}{1-n}\log\Tr(\chi^n). 
\end{aligned} \end{equation}

The first term corresponds to the R\'enyi entropy of the algebraic state, $S_{n}(\rho_a)$, while the second term is independent of the logical state. Moreover, since $\chi$ is a stabilizer state, its spectrum is flat, implying that its contribution is independent of $n$. This contribution defines the area term 
\begin{equation}\begin{aligned}
    \mathscr{A}=\frac{1}{1-n}\log\Tr(\chi^n). 
\end{aligned} \end{equation}

Since $\mathscr{A}$ doesn't depend on the R\'enyi index nor the logical information, it can be compute separately using a stabilizer reference state as before in the main text. Notice again that since the stabilizer part is concentrated to the area term, this piece returns the expected behavior of R\'enyi entropy for stabilizer states. Then the remaining portion of the R\'enyi entropy comes entirely from the bulk contribution $S_n(\rho_a)$ which we can compute by brute force. Since the bulk contribution captures all of the non-stabilizerness in the system, it is also the only source of non-flatness in the entanglement spectrum.

Knowing all R\'enyi entropies allows us to reconstruct the full entanglement spectrum of the target state $\ket{\Psi}$, which is given by
\begin{equation}
\mathrm{Spec}(\rho_A) 
= \bigcup_{\lambda \in \mathrm{Spec}(\rho_a)} 
\left\{ \underbrace{\tfrac{\lambda}{d_\chi}, \ldots, \tfrac{\lambda}{d_\chi}}_{d_\chi \text{ times}} \right\},
\end{equation}
where each $\lambda$ appears with degeneracy $d_\chi = \exp({\mathscr{A}})$.

% where $I$ is the identity matrix of dimension $d_{\chi}$ with $d_{\chi}=2^{\mathscr{A}}$. 

Based on the high degeneracy, one may alternatively measure a series of R\'enyi entanglement entropy with R\'enyi index from  $\alpha  = 2$ to $\alpha  = m$ through SWAP tests.  The logical space combined with the degeneracy gives $2^{\nu} + 1 = \text{poly}(n)$ degrees of freedom. Thus taking $m \sim \text{poly}(n)$ will be sufficient to generate enough $\exp( - (\alpha - 1) S_{\alpha} ) $ --- the power sums of the spectrum and invert it (by transforming to the elementary polynomial and solving the roots). However this requires an $\text{poly}(n)$ algorithm to compute $\exp( - (\alpha - 1) S_{\alpha} )$ where $\alpha \sim \text{poly}(n)$. A SWAP test approach with this complexity only works for $K \sim \mathcal{O}(1)$ for superpositions of stabilizers.

\section{A 5-qubit Example}\label{app:examples}

Here we provide an example of a 5-qubit state
\begin{equation}\begin{aligned}
\ket{\Psi}=&c_1\ket{s_1}+c_2\ket{s_2}+c_3\ket{s_3}+c_4\ket{s_4},
% \ket{s_1}:=&\left(\ket{00000}+\ket{10010}+\ket{01001}+\ket{10100}-\ket{11011}-\ket{11101}-\ket{00110}-\ket{01111}\right)\\
% \ket{s_2}:=&\left(\ket{00001}+\ket{10011}+\ket{01000}-\ket{10101}-\ket{11010}+\ket{00111}+\ket{11100}+\ket{01110}\right).
\end{aligned} \end{equation}
where $\ket{s_i}$'s are stabilized by a common stabilizer group $\mathcal{S}=\langle XZZXI,IXZZX,XIXZZ\rangle$.  $\ket{\psi}$ can be thought of as the codeword of the corresponding $[[5,2]]$ code. Take the first three qubits as a subsystem $A$.

We start with the binary symplectic representation of $\mS$: 
\begin{equation}
H_{S}=\left(
    \begin{array}{ccccc|ccccc|c}
        &&Z&& & &&X&&\\
        0 & 1 & 1 & 0 & 0 & 1& 0 &0 &1&0 &S_1\\
        0 & 0 & 1 & 1 & 0  &0& 1 &0 &0& 1& S_2\\
        0 & 0 & 0 & 1 & 1 & 1 & 0 & 1 & 0 & 0& S_3
    \end{array}
\right)
\end{equation}
In Step 1, we solve the kernel of the matrix to find the physical representative of the logical operators: 
\begin{equation}
H_L=\left(
    \begin{array}{ccccc|ccccc|c}
    &&Z&& & &&X&&\\
     0 & 1 & 1 & 0 & 0 & 0& 0 &0 &1&0 & \bar{X}_1\\
        0 & 0 & 1 & 0 & 0  &0& 0 &0 &0& 1 & \bar{X}_2\\
        1 & 0 & 1 & 1 & 0 & 0 & 0 & 0 & 0 & 0 &\bar{Z}_1\\
        0 & 1 & 0 & 0 & 1  &0& 0 &0 &0& 0 & \bar{Z}_2
    \end{array}
\right)
\end{equation}
The binary matrices satisfies ${H_S^X}{H_L^Z}^T+H_S^Z{H_L^X}^T=0 \mod 2$. 
Through row operations that corresponds to stabilizer multiplication, some of them can be written to have support entirely on $A$: 
\begin{equation}
H_{LA}=\left(
    \begin{array}{ccccc|ccccc|c }
    &&Z&& & &&X&&\\
    0 & 0 & 0 & 0 & 0 & 1& 0 &0 &0&0 & \bar{X}_1S_1\\
        1 & 0 & 1 & 0 & 0 & 0 & 1 & 0 & 0 & 0 &\bar{Z}_1\bar{X}_2S_2\\
        1 & 1 & 1 & 0 & 0  &1& 0 &1 &0& 0 & \bar{Z}_1\bar{Z}_2S_3
    \end{array}
\right)
\end{equation}
These three operators generate the algebra $\mathcal{M}_A$. Their physical representations are $XIIII, ZXZII, YZYII$. Through the decoding map they become $X_1, Z_1X_2, Z_1Z_2$ in the logical space. 

In Step 2, we reconstruct the logical state $\rho_a$ via, a $4\times 4$ matrix through
\begin{equation}\begin{aligned}\label{eq:reduced}
\rho_{a}&=2^{-k_a}\sum_{P\in \mathcal{M}_A} \bra{\psi}\bar{P}\ket{\psi}P&\\
    &=\frac{1}{4}\left(I\otimes I+
\Tr(\ket{\psi}\bra{\psi}XIIII)X\otimes I\right.\\
&\qquad \qquad \quad \,+\Tr(\ket{\psi}\bra{\psi}ZXZII)Z\otimes X \\
&\qquad \qquad \quad \,+\Tr(\ket{\psi}\bra{\psi}YZYII)Z\otimes Z  +\cdots),
\end{aligned} \end{equation}
and compute $S(\rho_a)$ by brute force. 

In Step 3, we find a reference state $|\phi \rangle$ stabilized by an extended set $\mS^{ext}=\langle XZZXI,IXZZX,XIXZZ, ZXIXZ,ZZZZZ\rangle$. The entanglement $S_{A}(\phi)=|A|-|\mathcal{S}_A^{ext}|=2\log 2$. The algebraic entropy $S(\phi_a)$ is $\log 2$. Therefore $\mathscr{A} = \log 2$. We conclude that $S_A( \psi) = S(\rho_a(\psi) ) + \log 2$.

\section{Extracting the maximal stabilizer group }\label{app:stablearn}
\subsection{Input state as superposition of stabilizer states}\label{app:superposition}
When the target state can be decomposed as a superposition of a small number of stabilizer states, our algorithm is particularly effective if these states share a sufficient number of common stabilizers. Even when this condition is not fully met, knowledge of the stabilizer decomposition can still be useful, as it may reveal additional elements of the stabilizer group of the target state.

It is not unusual to encounter a collection of states $\{|\psi_j\rangle\}$ that are already common eigenstates of certain Pauli operators, possibly with opposite eigenvalues, (for instance, wavefunction written in the computational basis). In this case, standard methods \cite{Gottesman1998TheHR} can quickly determine their common stabilizer group. Here we would like to investigate more general sets of stabilizer states and identifying their common stabilizer group. 

It is important to note that different signs for the eigenvalues correspond to different stabilizers. For example, $|s_1 \rangle = |0 \rangle$, $|s_2 \rangle = |1 \rangle $. The stabilizers are $Z$, but with eigenvalues $+1$ and $-1$, respectively. Thus their stabilizer groups are generated by $Z$ and $-Z$, which are distinct. When we form the superposition $|0\rangle + |1\rangle$, the stabilizer switches to $X$, rather than $Z$.

We begin by detailing the procedure for identifying stabilizers of states that can be expressed as a superposition of two stabilizer states,
$|\psi\rangle=a|s_1\rangle+b|s_2\rangle$, where $|s_1\rangle$ and $|s_2\rangle$ are stabilizer states with stabilizer groups $\mathcal{S}_1$ and $\mathcal{S}_2$, respectively. To determine the full set of stabilizers of $|\psi\rangle$, we proceed as follows:
%{\color{red} case $|s_1\rangle, |s_2\rangle$ orthogonal}
 % In the following, we treat each stabilizer as a binary vector and the stabilizer group as a vector space. First step we find a complete basis for the vector space  $\mathcal{S}_0=\mathcal{S}_1\cap \mathcal{S}_2$, and rewrite the two stabilizer groups as $\mathcal{S}_1=\mathcal{S}_0\oplus \mathcal{S}_1'$, $\mathcal{S}_2=\mathcal{S}_0\oplus \mathcal{S}_2'$. Such that the two reduced stabilizer groups $\mS_1'$ and $\mS_2'$ are totally independent. 
  
 \begin{enumerate}
    \item Compute the intersection of the two stabilizer groups,  
\[
\mathcal{S}_0 = \mathcal{S}_1 \cap \mathcal{S}_2,
\]  
which consists of the stabilizers common to both groups. It suffices to identify a generating set for this intersection. For simplicity, we restrict our attention to the case where all elements of $\mathcal{S}_0$ appear with the same sign in both $\mathcal{S}_1$ and $\mathcal{S}_2$. 

% {\color{blue} When both group generators are given in terms of their check matrices $H_1, H_2$, then $\mathcal{S}_0$ can be identified by $ker(H_1)\cap ker(H_2)$, which one can do by stacking the matrices and performing Gaussian elimination.} 

    \item Decompose each stabilizer group $\mS_1$ (or $\mS_2$) into its shared part $\mS_0$, and unique part, $\mS_1'$ (or $\mS_2'$):
    \[
    \mathcal{S}_1 = \langle \mathcal{S}_0, \mathcal{S}_1' \rangle, 
    \qquad 
    \mathcal{S}_2 = \langle \mathcal{S}_0, \mathcal{S}_2' \rangle,
    \]  
    where $\mathcal{S}_1'$ and $\mathcal{S}_2'$ are independent ($\mS_1'\cap\mS_2'=I$).  

    \item Expand $\ket{\psi}\bra{\psi}$ in terms of the group elements drawn from $\mathcal{S}_0$, $\mathcal{S}_1'$, and $\mathcal{S}_2'$.  
    \begin{equation}\begin{aligned}\label{eq:stateexp}
    \ket{\psi}\bra{\psi} 
    =&\prod_{k=1}^{n-r}\frac{1+S_0^k}{2}\left(|a|^2\prod_{i=1}^r\frac{I+S_1'^i}{2}+|b|^2\prod_{j=1}^r\frac{I+S_2'^j}{2}\right.\\
    &+\left.\frac{ab^*}{\bra{s_1}s_2\rangle}\prod_{i=1}^r\frac{I+S_1'^i}{2}\prod_{j=1}^r\frac{I+S_2'^j}{2}+h.c.\right).
\end{aligned} \end{equation}

    \item Classify each Pauli operator according to whether it belongs to $\mathcal{S}_1'$, $\mathcal{S}_2'$, or the group generated by $\langle \mathcal{S}_1' \cup \mathcal{S}_2' \rangle$. The Pauli coefficients are then determined according to this classification.  
    \item Identify all Pauli operators in the expansion whose coefficients have unit norm. Together with $\mS_0$, these operators form the complete stabilizer group of $\ket{\psi}$.   
\end{enumerate}
 
%{\color{red}sry what's $|\psi'\rangle$ supposed to be?}
To illustrate the last two steps, we define $\ket{\psi'}$ to be the logical state of $\ket{\psi}$ with respect to the code defined by stabilizer group $\mS_0$. so that,   
\begin{equation}
    \ket{\psi}\bra{\psi} \;=\; 
    \prod_{k=1}^{n-r} \frac{1+S_0^k}{2}\, \ket{\psi'}\bra{\psi'}.
\end{equation}
from the state expansion in Eq.~\eqref{eq:stateexp}. 

The Pauli coefficients of $\ket{\psi'}$ can be written explicitly as follows:  
\begin{enumerate}
    \item $\bra{\psi'}P\ket{\psi'}=|a|^2+2^{1-r}\Re{\frac{a^*b}{\bra{s_1}s_2\rangle}}$ for $P=g_1^i$ belong to $\mS_1'$.
    \item $\bra{\psi'}P\ket{\psi'}=|b|^2+2^{1-r}\Re{\frac{a^*b}{\bra{s_1}s_2\rangle}}$ for $P=g_2^i$ belong to $\mS_2'$.
    \item $\bra{\psi'}P\ket{\psi'}=2^{1-r}\Re{\frac{a^*b}{\bra{s_1}s_2\rangle}}$ for $P=g_1^ig_2^j$, if $g_1^ig_2^j=g_2^jg_1^i$
    \item $\bra{\psi'}P\ket{\psi'}=2^{1-r}\Im{\frac{a^*b}{\bra{s_1}s_2\rangle}}$ for $P=g_1^ig_2^j$, if $g_1^ig_2^j=-g_2^jg_1^i$
    
\end{enumerate}
     Thus, it suffices to check these four coefficients: if any of them has unit norm, the corresponding Pauli operator is a stabilizer of $\ket{\psi}$.
     
\subsection{Finding the maximal stabilizer group for more general cases}\label{app:generalcase}
The situation is more complicated when some of the stabilizers of $\ket{s_1}$ and $\ket{s_2}$ have identical Pauli type but different signs, i.e. they are orthogonal. We denote the sign-stripped version of these stabilizers as $\mS_0'$.  In this case the overlap $\langle s_1\ket{s_2}=0$. So we can't directly apply the expansion as in the previous section. The strategy is to insert error operators $\{E_k\}$, each of which only anti-commute with one element of $S_0'$. More specifically, we find set of $\{E_k\}$, such that $\{E_k,S_0'^k\}=0$, and $[E_k,S_0'^{l\neq k}]=0$. They also commute with other stabilizers in $\mS_1$ and $\mS_2$, so $[E_k,S_1']=0$, $[E_k,S_2']=0$. Then we can write the reduced state as 
{\small
\begin{align}
    &\ket{\psi'}\bra{\psi'} \\\nonumber
    =&\Big(|a|^2\prod_k\frac{1+S_0'^k}{2}\prod_{i}\frac{I+S_1'^i}{2}+|b|^2\prod_k\frac{1-S_0'^k}{2}\prod_j\frac{I+S_2'^j}{2}\\\nonumber
    &+\frac{ab^*}{\bra{s_1}\prod_kE_k\ket{s_2}}\prod_k\frac{1+S_0'^k}{2}\prod_i\frac{I+S_1'^i}{2}\prod_j\frac{I+S_2'^j}{2}\prod_kE_k\\\nonumber
    &+h.c.\Big)
\end{align}
}%

Note that $E:=\prod_kE_k$ does not belong to  $\mS_0'\oplus\mS_1'\oplus\mS_2'$, and hence independent from any element in the joint group. The non-zero expectation values can be classified as  
\begin{enumerate}
    \item $\bra{\psi'}P\ket{\psi'}=(|a|^2\pm |b|^2)$ for $P=g_0^i$ generated by elements in $\mS_0'$. 
    \item $\bra{\psi'}P\ket{\psi'}=|a|^2$ for $P=g_0^ig_1^j$, and $g_1^j\neq I$.
    \item $\bra{\psi'}P\ket{\psi'}=|b|^2$ for $P=g_0^ig_2^j$, and $g_2^j\neq I$.
    \item $\bra{\psi'}P\ket{\psi'}=2^{1-r_1}\Re\left(\frac{ab^*}{\bra{s_2}E\ket{s_1}}\right)$ for $P= g_0^ig_1^jg_2^kE$, if $g_0^ig_1^jg_2^kE=Eg_2^kg_1^jg_0^i$.
    \item $\bra{\psi'}P\ket{\psi'}=2^{1-r_1}\Im\left(\frac{ab^*}{\bra{s_2}E\ket{s_1}}\right)$, for $ P=g_0^ig_1^jg_2^kE$, if $g_0^ig_1^jg_2^kE=-Eg_2^kg_1^jg_0^i$.
\end{enumerate}

% {\color{red}idk what this sentence is supposed to say}
As in the previous section,  we only need to check few coefficients to find the stabilizers whose expectation values is $\pm 1$.  

Now consider a linear superposition of $K$ stabilizer states $\{\ket{s_i},i=1,\dots,K\}$ such that, 
\begin{equation}\begin{aligned}
    \ket{\psi}=\sum_{i=1}^Kc_i\ket{s_i}. 
\end{aligned} \end{equation}
The density matrix of this state can be expanded as, 
\begin{equation}\begin{aligned}
    \ket{\psi}\bra{\psi}=&\sum_{i=1}^K|c_i|^2\prod_{a=1}^n\frac{I+S_{i}^a}{2}\\
    &+\sum_{i\neq j}\frac{c_i^*c_j}{\bra{s_i}E_{ij}\ket{s_j}}\prod_{a=1}^n\frac{I+S_{i}^a}{2}E_{ij}\prod_{b=1}^n\frac{I+S_{j}^b}{2}
\end{aligned} \end{equation}

For each stabilizer group, we associate a binary vector space $V_i$. The product of stabilizers is represented by the direct sum $V_i \oplus V_j$. There is an error operator $E_{ij}$ inserted when a pair of states are orthorgonal, $\langle s_i\ket{s_j}=0$. The corresponding vector is denoted as $\vec e_{ij}$. Given a $2n$-dimensional binary vector $\vec{v}$, we construct a truth vector of length $K(K+1)/2$ that records whether $\vec{v} \in V_i$ and $\vec{v}+\vec{e}_{ij}\in V_i \oplus V_j$ for all $i,j$. Using this truth vector, the corresponding Pauli spectrum is obtained by summing the coefficients (up to sign) of terms with truth value $1$. The total number of distinct truth vectors scales as $2^{K^2/2}$, which gives an upper bound on the complexity of determining both the full Pauli spectrum and the stabilizers of $|\psi\rangle$.

\subsection{Quantum algorithm for learning stabilizers}\label{app:quantumlearn}

When the input is an unknown quantum state $\ket{\Psi}$, one can find its maximal stabilizer group using the algorithm of \cite{Grewal:2023hzn}, which learns $\mathrm{STAB}(\ket\Psi)$ in $\mathrm{poly}(n)$ time with $\mathrm{poly}(n)$ copies of the state. Combining this with our method yields an efficient quantum algorithm for learning the entanglement entropy of low-magic states.

% The algorithm is outlined as follows: 

This algorithm is based on \textit{Bell Difference Sampling} \cite{Montanaro:2017oht}. Let $\ket{\Psi}$ be a state of  $n$ qubits,  the bell sampling is defined as the following procedure: 

\begin{enumerate}
    \item Create two copies of $\ket{\Psi}$. Label their qubits as $A_1, A_2,\cdots A_n$, and $B_1,B_2,\cdots B_n$.
    \item Measure each pair $A_iB_i$ in the bell basis:
    \begin{equation}
        \frac{\ket{00}+\ket{11}}{\sqrt 2}, \quad \frac{\ket{00}-\ket{11}}{\sqrt 2}, \quad \frac{\ket{01}+\ket{10}}{\sqrt 2}, \quad \frac{\ket{01}-\ket{10}}{\sqrt 2}
    \end{equation}and obtain a bit string of length $2n$. 
    \item Repeating this process to obtain a set of measurement results $\{r_0, r_1,\cdots r_m\}$, where each $r_i$ is a $2n$ length bit-string. 
    \item Create a new set $S=\{r_1\oplus r_0, r_2\oplus r_0,\cdots, r_m\oplus r_0\}$. Determine a basis $B$ for $S$ in the linear space $\mathbb{F}_{2}^{2n}$. 
    \item With high probability (determined by the number of samples $m$) the set $B$ corresponds to the commutant of $\stab(\Psi)$, denoted as $\stab(\Psi)^{\perp}$.
\end{enumerate}

 Let $\ket{\psi}$ be a state of $n$ qubits. The Bell Sampling on two copies of $\psi$'s returns outcome $r$ with probability, 
    \begin{equation}
        p_{\psi}(r)=\frac{|\langle\psi|\sigma_r\ket{\psi^*}|^2}{2^n}.
    \end{equation}

Therefore by sampling the bell measurement in a quantum computer, we're able to learn the support of this distribution. This will determine the stabilizer group of target state due to the following theorem: 

\begin{theorem}
    Let $\mathcal{S}=\stab(\Psi)$ being the stabilizer group of state $\ket{\Psi}$, and  $\N(\mathcal{S})=\stab{(\Psi)}^{\perp}$ being the normalizer group of $\mathcal{S}$. Then the support of Bell sampling lies in $\N(\mathcal{S})$.
\end{theorem}

This is proved in \cite{Grewal:2023hzn} through the duality equation. Here we provide another proof based on the QECC property:

\begin{proof}
    The stabilizer group $\mathcal{S}=\stab(\Psi)$ defines a stabilizer QECC, where $\ket{\Psi}$ is in the code subspace $\mathcal{H}_{code}$. All the Pauli operators are either belong to the normalizer $\N(\mathcal{S})$ if it commute with all elements in $\mathcal{S}$, or it belongs to the set of error operators, which anti-commute with some elements in $\mathcal{S}$. Any error operator $E$ must have zero expectation value in the code subspace, because for any two states $\ket{\phi},\ket{\psi}\in \mathcal{H}_{code}$, we have 
    \begin{equation}
        \bra{\phi}E\ket{\psi}=\bra{\phi}\{E,S\}\ket{\psi}, \qquad \text{for $\forall S\in \mathcal{S}$}
    \end{equation}
    since they are eigenstates of $\mathcal{S}$ with $+1$ eigenvalue. But there are some $S\in \mathcal{S}$ anti-commute with $E$. Therefore these matrix elements must vanish. 

    Since the support of Bell sampling must have non-zero expectation value, we conclude that they must belong to $\N(\mathcal{S})$. 
\end{proof}

% Then we show that the Pauli group generated by the elements in the support must be the full normalizer group. Otherwise, denote the group generated by the support as $G$ and $G\subsetneq \N(\mathcal{S})$. 

% For $G=\stab(\psi)$, this equation indicates that $p_{\psi}$ has support lies in $\stab{(\psi)}^{\perp}$,
% \begin{equation}
%     \sum_{r\in \stab(\psi)^{\perp}}p_{\psi}(r)=1.
% \end{equation}

After finding $\N(\mathcal{S})$, there is a classical algorithm to find $\mathcal{S}=\stab(\Psi)$ in $\mathcal O(n^3)$ time. Putting together, the effectiveness of this quantum algorithm is summarized in the following theorem.

% this is an efficient quantum algorithm which takes $\poly(n)$ resources and learns stabilizers of the state $\ket{\Psi}$, approximately as summarized in the following theorem. 

\begin{theorem}
    ( Theorem 7.1 of \cite{Grewal:2023hzn}): with success probability of $1-\delta$. It takes $\mathcal O\left(\frac{32\log(1/\delta)+64n}{\epsilon^2}\right)$ copies of the state $\ket{\Psi}$, and time of $\mathcal O\left(\frac{n^2\log(1/\delta)+n^3}{\epsilon^2}\right)$ to measure the stabilizer group $G$ of a state $\ket{\hat\Psi}$, which is $\epsilon$ close to $\ket{\Psi}$, and has a non-smaller stabilizer group, i.e. $\stab(\hat\Psi)\supseteq \stab(\Psi)$ and $|\Psi-\hat\Psi|<\epsilon$.
\end{theorem}

% More specifically, from Theorem 7.1 of \cite{Grewal:2023hzn}, there is an algorithm with success probablity of $1-\delta$. It takes $O\left(\frac{32\log(1/\delta)+64n}{\epsilon^2}\right)$ copies of the state $\ket{\psi}$, and time of $O\left(\frac{n^2\log(1/\delta)+n^3}{\epsilon^2}\right)$ to measure the stabilizer group $G$ of a state $\ket{\hat\psi}$, which is $\epsilon$ close to $\ket{\psi}$, and has a non-smaller stabilizer group, i.e. $|\stab(\hat\psi)|\geq |\stab(\psi)|$ and $|\psi-\hat\psi|<\epsilon$.

% The Bell Sampling algorithm is based on the following Lemma \cite{Montanaro:2017oht}:
% \begin{lemma}
%     Let $\ket{\psi}$ be a state of $n$ qubits. The Bell Sampling on two copies of $\psi$'s returns outcome $r$ with probability, 
%     \begin{equation}
%         p_{\psi}(r)=\frac{|\langle\psi|\sigma_r\ket{\psi^*}|^2}{2^n}.
%     \end{equation}
% \end{lemma}

% Another important ingredient is the duality  relation \cite{Grewal:2023hzn},
% \begin{equation}
%     \sum_{a\in G} p_{\psi}(a)=\frac{|G|}{2^n}\sum_{a\in G^{\perp}}p_{\psi}(a).
% \end{equation}

Then we run the algorithm~\ref{alg:combined} to find the area entropy as well as the logical operators supported on subregion $A$, and perform tomography for the algebraic bulk state, which has complexity of $e^\nu$.

This algorithm finds the stabilizer group of the state $\ket{\hat\Psi}$ that is $\epsilon$ close to the target state $\ket{\Psi}$.   Then by Audenaert-Fannes-Petz inequility \cite{Fannes1973,Audenaert_2007}, their Von Neumann entropy are bounded by 
\begin{equation}
    |S_A(\hat\Psi)-S_A(\Psi)|\leq |\hat\Psi-\Psi|\log(d-1)+H(|\hat\Psi-\Psi|)
\end{equation}
where $H(T)=-T\log T-(1-T)\log (1-T)$ is the Shannon entropy of the trace distance. So we conclude that, to bound the entropy error by $\epsilon$, we need the trace distance to be $\epsilon/n$. The time complexity and sampling complexity are still $\poly(n)$ scaling. 

Based on this analysis, we define a smoothed version of stabilizer nullity, 
\begin{equation}\begin{aligned}
    \nu^{\epsilon}(\Psi):=\min_{\hat\Psi \in \mathcal{B}^{\epsilon}(\Psi)} \nu(\hat\Psi).
\end{aligned} \end{equation}
It is the smoothed stabilizer nullity that determines the complexity of calculating entanglement entropy up to error of $\mathcal O(n\epsilon)$.

The smoothed stabilizer nullity can be bounded by the continuous \textit{Stabilizer Renyi Entropy} (SRE) function \cite{PhysRevLett.128.050402} for $\alpha<1$ as follows, using the inequility in \cite{1365269}:
\begin{equation}\begin{aligned}
    M_{\alpha}(\Psi)\leq \nu^{\epsilon}(\Psi)\leq M_{\alpha}(\Psi)+\frac{1}{1-\alpha}\log{1/\epsilon}.
\end{aligned} \end{equation}

\section{Applications of the algorithm }\label{app:application}

\subsection{Clifford circuit + few non-Clifford gates}

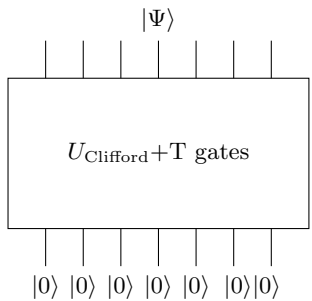
\begin{figure}
    \centering
    \begin{tikzpicture}
        \draw[draw=black] (0,0) rectangle ++(4,2);
        \draw (2,1) node {$U_{\text{Clifford}}$+T gates};
        \draw (0.5,0)--(0.5,-0.5) node[below]{$\ket{0}$};
        \draw (1,0)--(1,-0.5) node[below]{$\ket{0}$};
        \draw (1.5,0)--(1.5,-0.5) node[below]{$\ket{0}$};
        \draw (2,0)--(2,-0.5) node[below]{$\ket{0}$};
        \draw (2.5,0)--(2.5,-0.5) node[below]{$\ket{0}$};
        \draw (3,0)--(3,-0.5);
        \draw (3.25,-0.5) node[below] {$\ket{0}\ket{0}$};
        \draw (3.5,0)--(3.5,-0.5);
        \draw (0.5,2)--(0.5,2.5) ;
        \draw (1,2)--(1,2.5);
        \draw (1.5,2)--(1.5,2.5);
        \draw (2,2)--(2,2.5) node[above]{$\ket{\Psi}$};
        \draw (2.5,2)--(2.5,2.5) ;
        \draw (3,2)--(3,2.5);
        \draw (3.5,2)--(3.5,2.5);
    \end{tikzpicture}
    \caption{Clifford circuit with few number of non-Clifford single qubits gates, acting on initial product state.   }
    \label{fig:TC}.
\end{figure}

In the first situation we consider Clifford circuit doped with $t$ number of non-Clifford single-qubit gates. Shown in Fig.~\ref{fig:TC}. The statbilizer nullity of this state is upper bounded by $2t$ according to the following theorem:

\begin{theorem}
    \cite{10.1145/3618260.3649738}: Let $\ket{\Psi}$ be produced by Clifford circuit doped by $t$ number of single-qubit non-Clifford gates, then $|\stab(\ket{\Psi})|\geq 2^{n-2t}$, i.e. the stabilizer nullity is at most $2t$.
\end{theorem}

We compute the entanglement entropy of $\ket{\Psi}$ by following the steps outlined in Algorithm~\ref{alg:combined}. The key tasks are to identify the stabilizer group of $\ket{\Psi}$ and then compute the expectation values of the logical operators, which enables state tomography of the logical subsystem. 

To determine the stabilizer group, we evolve the Pauli group generated at the bottom layer by $\mathcal{S}_{0}=\langle Z_1,Z_2,\cdots Z_n\rangle $ through the Clifford circuit. Whenever a non-Clifford single-qubit gate is encountered, we perform Gaussian elimination to obtain a new generating set that avoids overlap with the non-Clifford gate. At the output layer, this procedure yields the stabilizer group $\mathcal{S} = \mathrm{STAB}(\ket{\Psi})$, which is generated by at least $n - 2t$ Pauli operators.

Next, we evaluate the expectation values of logical operators. After identifying the logical operators supported on the subregion $A$, which form a group $G_A$, we evolve them backward through the circuit. Each time a non-Clifford gate is encountered, a logical operator branches into a superposition of at most four Pauli strings. 

At the bottom layer, the backward-evolved logical group is denoted by $G_{A}^{\mathcal{C}}:=\mathcal{C}^{\dagger}G_{A}\mathcal{C}$, where each element is a superposition of at most $4^t$ Pauli operators. We then compute the expectation values of all elements in $G_A^{\mathcal{C}}$ with respect to the input tensor-product state $\ket{\phi_0} = \ket{0}^{\otimes n}$. 
\begin{equation}
    \bra{\Psi}P\ket{\Psi}=\bra{\phi_0}\mathcal{C}^{\dagger} P\mathcal{C}\ket{\phi_0}, \quad \text{for $P$ in $G_A$.}
\end{equation}

% \begin{equation}
% \rho_a = \sum_{g_a \in G_A^{\mathcal{C}}} \Tr(\phi_0 \bar{g}_a) g_a.
% \end{equation}

The cost of evaluating each expectation value is determined by the number of branches, which is upper-bounded by $4^t$. Since $G_A$ contains at most $4^{2t}$ elements, the total computational complexity of obtaining entanglement entropy scales as $\mathcal O(4^{3t})$.

\subsection{Magic initial state + Clifford circuit }

Similar complexity bound of  entropy calculation for state prepared by Clifford circuit acting on non-stabilizer initial state can also be derived. Suppose $t$ qubits of the initial layer are used to prepare a magic state. 

\begin{figure}
    \centering
    \begin{tikzpicture}
        \draw[draw=black] (0,0) rectangle ++(4,2);
        \draw (2,1) node {$U_{\text{Clifford}}$};
        \draw (0.5,0)--(0.5,-0.5) node[below]{$\ket{0}$};
        \draw (1,0)--(1,-0.5) node[below]{$\ket{0}$};
        \draw (1.5,0)--(1.5,-0.5) node[below]{$\ket{0}$};
        \draw (2,0)--(2,-0.5) node[below]{$\ket{0}$};
        \draw (2.5,0)--(2.5,-0.5) node[below]{$\ket{0}$};
        \draw (3,0)--(3,-0.5);
        \draw (3.25,-0.5) node[below] {$\ket{\text{magic}}$};
        \draw (3.5,0)--(3.5,-0.5);
        \draw (0.5,2)--(0.5,2.5) ;
        \draw (1,2)--(1,2.5);
        \draw (1.5,2)--(1.5,2.5);
        \draw (2,2)--(2,2.5) node[above]{$\ket{\psi}$};
        \draw (2.5,2)--(2.5,2.5) ;
        \draw (3,2)--(3,2.5);
        \draw (3.5,2)--(3.5,2.5);
    \end{tikzpicture}
    \caption{Clifford acting on initial tensor product of qubits, with $t$-number of magic states.   }
    \label{fig:TC2}.
\end{figure}
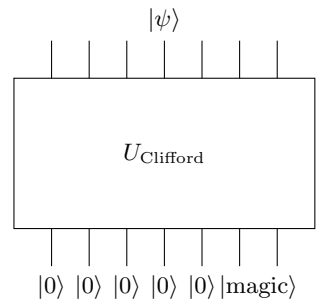

Then the worst case complexity is when this state has stabilizer group of rank $n-t$. The number of logical operators supported on subregion $A$ is at most $4^t$. Calculating the expectation value of each logical operator has complexity  $\mathcal O(4^t)$. So the total complexity is bounded by $\mathcal O(4^{2t})$.

%\subsection{Magic augmented Clifford circuits}
%\CC{Add more about high magic high entanglement}
However, one can sometimes avoid the exponential scaling with nullity if we are given further structure of the magic state.  For instance, when the input state is a product state, then there are states for which the algorithm remains efficient even when the amount of magic is $\mathcal O(n)$. One example we have given was the holographic code, where the Clifford unitary is then given by the encoding unitary of the code where the size of product magic states $k$ is equal to the number of bulk or logical qubits in the code.

Such states are special instances prepared by magic-augmented Clifford circuits. Consider states prepared by circuits acting initial $|0\rangle$ states where the unitary circuit consists of single-qubit non-Clifford gates which precede and/or follow  the Clifford circuits \cite{Zhang:2025dhg}. Since the entanglement spectrum is invariant under local unitary rotations, it suffices to consider the class of states prepared by circuits where all single-qubit non-Clifford gates happen before the Clifford circuit. 

For each such circuit, one can associate a stabilizer code for which the Clifford is the encoding circuit where both the stabilizer group and the logical representations of the $k$ qubits are known. The number of magical states $k$ determines the size of the code subspace whereas the remaining $n-k$ $|0\rangle$ states are the so-called syndrome bits \cite{Ferris_2014} which we keep fixed. Note that these syndrome bits are not to be confused with the ancilla qubits used in syndrome extraction circuits. Rather, if $U_{\rm Clifford}^{\dagger}$ is applied to a state in an error subspace of a definite syndrome, then one or more of the syndrome bits will be in $|1\rangle$ as opposed to $|0\rangle$ states.

For each such code, one can efficiently characterize the logical Pauli subalgebra through binary matrix operations where one can determine how much information of the $k$ logical qubits are recoverable from subregion $A$ where $S(A)$ can again be computed with the generalized RT formula on stabilizer codes. To see how this can work, first consider a simple example where the code satisfies complementary subsystem error correction, i.e. if $\mathcal M_A$ has a trivial center, where the subregion $A$ recovers precisely $k'$ of the $k$ logical qubits of information. Then by construction the total bulk entropy must be zero and one can then compute the area contribution efficiently by setting the $k$ qubits to product of $|0\rangle$ and computing $S(A)_{U_{\rm Clifford}|0\rangle^{\otimes n}}$. This hold regardless of $k$ or $S(A)$, both of which can scale linearly with the system size. Such is the case of subsystem holographic stabilizer codes, where all bulk qubits in the ``entanglement wedge'' of $A$ are exactly recoverable. Then regardless of the states of the bulk qubits, as long as they are not entangled with those in the complementary wedge, the total entanglement of the system is given by the area of the minimal surface, i.e. the minimum number of edge cuts that separates $A$ and $A^c$ in the tensor network. An example of this is shown in Fig.~\ref{fig:holo}, but the conclusion holds for higher dimensional codes also where the area of the minimal surface, to leading order, will scale the same way as the area of $|\partial A| \sim n^{1-1/D}$. Generalizing, if $U_{\rm Clifford}$ is an encoding unitary where there exists a recovery map restricted to $A$ such that subsets of the logical qubits are recoverable, then the entanglement of $S(A)$ is efficiently computable.

Generally, however, a stabilizer code only satisfies subalgebra complementary error correction. This means that the Pauli subalgebra supported on $A$ need not correspond to full qubits. Instead, the recoverable Pauli operators can always be generated by some combinations of anticommuting pairs $\mathcal{PF}_A=\{({P}_i,{Q}_i), i=1,\dots, k'\}$ where $\{{P}_i,{Q}_i\}=0$ which pairwise commute and lone Pauli operators $\{{R}_j\}=Z_M$ that commutes with all other logical operators supported on $A$ and form the center of $\mathcal M_A$. It follows from the properties of symplectic vector spaces \cite{artin2016geometric} that any Pauli subalgebra can be written in a basis of the above form. Importantly, $({P}_i$, ${Q}_i)$ and ${R}_i$ generally do not implement the canonical logical $(X_i, Z_i)$ and $X_i$ (resp. $Y_i, Z_i$) of the encoded logical qubit. Instead, it could have support over many logical qubits. An example of this shown in Appendix~\ref{app:examples} where the anticommuting pair can be written as $Y_1X_2, Y_1Z_2$ while the center element is $X_1Y_2$.

If $(X_i, Z_i)\in \mathcal{PF}_A$, then the corresponding logical state $|\psi_i\rangle$ must be recoverable by \cite{Harlow2016TheRF}. By assumption the state is pure and does not contribute to bulk entropy. In a similar way, if the canonical logical Pauli $R_i$ ($R=X,Y$ or $Z$) is recoverable on $A$, then a classical bit $\Pi_R |\psi_i\rangle\langle \psi_i|\Pi_R$  is recoverable where $\Pi_R$ is the projection onto the Pauli algebra generated by $R$. In this case, $S(\Pi_R |\psi_i\rangle\langle \psi_i|\Pi_R)$ returns its bulk entropy contribution. The bulk entropies for these qubits can then be added one by one because the logical state is a tensor product.

Then let the Pauli subalgebra generated by the remaining recoverable Pauli that cannot be written in the canonical basis of logical Paulis be $\mathcal{G}_A$. We will refer to it as the non-canonical logical algebra. The bulk entropy from these terms is generally non-trivial $S(\rho_a;\mathcal{G}_A) = S(\Pi_{\mathcal{G_A}}\rho_a\Pi_{\mathcal{G_A}})$ where $\Pi_{\mathcal{G_A}}$ is the projection onto the subalgebra $\mathcal{G}_A$. Because the Pauli operators here have support over many qubits, one cannot simply add the entropies term by term like above. Hence it has to be computed by brute force, leading to a complexity that is $\mathcal O({|\mathcal{G}_A|})$. Adding up the bulk entropies from above from both the canonical and the non-canonical parts, we obtain the total bulk entropy, which is the most costly part of the computation. The area term can be computed like before by choosing a logical stabilizer reference state. Therefore, the size of the ``non-canonical'' logical algebra $\mathcal G_A$ sets a lower bound of the computational complexity of our algorithm. %The best possible scenario is when all anticommuting $(X_i,Z_i)$ pairs and center elements in the original basis are recoverable on $A$. 

%One can identify $U_{\rm Clifford}$ for which $\mathcal{G}_A$ is small in a number of ways, but typical concatenated codes and convolutional codes where the seeds are good erasure correction codes will typically result in $\mathcal{G}_A=O(1)$ for bipartitions where either $A$ or $A^c$ is a correctable erasure.

% and this lower bound is saturated by some state in the Clifford orbit of the input product magic states because there is always a global Clifford change of basis over the logical subspace that can reorganize the anti-commuting pairs of Pauli operators into tensor product of qubits.

\bibliographystyle{apsrev4-1}
\bibliography{ref}

\end{document}